\documentclass[12pt]{article}

\usepackage[linktocpage=true,plainpages=false]{hyperref}
\usepackage{color}
\usepackage{amsmath,amssymb}
\usepackage{cite}
\usepackage{array}
\usepackage[T1]{fontenc}
\usepackage[labelsep=period]{caption}

\definecolor{linkcolor}{rgb}{0.6,0,0}
\definecolor{citecolor}{rgb}{0,0.6,0}
\definecolor{urlcolor}{rgb}{0,0,0.9}
\hypersetup{colorlinks, linkcolor={linkcolor},citecolor={citecolor}, urlcolor={urlcolor}}
\begin{document}
\title{Field-Theoretical Formulation\\ of Regge-Teitelboim Gravity}
\author{A. A. Sheykin\thanks{E-mail: a.sheykin@spbu.ru},
S. A. Paston\thanks{E-mail: s.paston@spbu.ru}\\
{\it Saint Petersburg State University, Saint Petersburg, Russia}
}
\date{\vskip 15mm}
\maketitle

\begin{abstract}
Theory of gravity is considered in the Regge-Teitelboim approach in which the pseudo-Riemannian space is treated as a surface isometrically embedded in an ambient Minkowski space of higher dimension. This approach is formulated in terms of a field theory in which the original pseudo-Rimannian space is
defined by the field constant-value surfaces. The symmetry properties of the proposed theory are investigated, and possible structure of the field-theoretical Lagrangian is discussed.
\end{abstract}

\newpage

\section{Introduction}
Formulating a consistent theory of gravity is one of
the priorities of contemporary theoretical physics.
Einstein's general relativity (GR) is supported by a
large number of astrophysical observations, but cannot claim to be the ultimate gravitational theory. Primarily, this is because none of the many attempts to quantize this theory has fully succeeded \cite{carlip}. On the other hand, the GR predictions are at odds with some observational data, and one has to introduce extra entities like "dark matter" for describing the galactic rotation curves and "dark energy" for interpreting the rate of expansion of the Universe \cite{gorbrub}. Therefore,
the general relativity has to be extended so as to allow
for a consistent quantization and/or to describe the
aforementioned phenomena.

One of such extensions is the theory proposed in
1975 by Regge and Teitelboim \cite{regge}. In this string-inspired approach
the theory of gravity is treated as the theory of a four-dimensional surface embedded in an ambient ten-dimensional Minkowski space. A major asset of this approach is that the ambient space is flat in contrast with the brane theory which
attracted much attention in the last years. Indeed, the
quantization of gravity is complicated by the "background problem": no canonical quantization is possible on the Minkowski space as a background, whereas
the path-integral quantization is plagued by problems
with defining the measure. That the ambient Minkowski space is not observable in the Regge-Teitelboim approach allows one to circumvent the background problem and to formulate the concepts required for quantization (such as the preferred time direction and the definition of causality). Apart from this, the dynamic equations of the theory (known as Regge-Teitelboim equations) are more general than Einstein
ones and possibly could provide a clue to the concepts
of dark matter and dark energy.

Despite some definite advantages with respect to
the conventional GR, the considered approach inherits a number of problems mostly associated with introducing the coordinates $x^{\mu}$ on the surface in the ambient space defined by the embedding function. In order
to do away with the coordinates in the embedding the
ory, it has been proposed \cite{statja25} to introduce the field
$z^A(y^a)$ ($A=0\ldots5$) in the ambient space $y^a$, $a=0\ldots9$. Then, the constant-value surfaces
$z^A(y^a)=\text{\text{const}}$ (further denoted as $\mathcal{M}$) are four-dimensional surfaces
in the embedding space that may be defined without
introducing the coordinates. This transition results in
a field theory in flat spacetime that may be reduced
to the Einstein theory  \cite{statja25}. Rather than require the
equivalence to GR, in this paper, we consider a viable
form of the Lagrangian in the field-theoretical framework. Upon briefly formulating the discussed splitting
theory \cite{statja25}, we proceed to probe the Lagrangian form
and to compare the predictions with those of GR.

\section{Scalars of the splitting theory}
Let us consider a possible Lagrangian form in the
splitting theory which involves a set of scalar fields $z^A (y^a)$ defined in the ten-dimensional Minkowski space. An
important transformation of the theory is the surface
renumbering, which is analogous to diffeomorphism,
\begin{align} \label{1}
	z^A \to z'^A \equiv f^A(z^B),
\end{align}
where $f$ is an arbitrary function. Since the observable
quantities should depend on the surface properties
rather than the numbering order, the theory should be
invariant with respect to this transformation. Let us
consider the invariants under this transformation that
may be constructed using the variables of the theory.

Since according to  (\ref{1}) no nontrivial invariants may
be constructed using the variables $z^A$, the Lagrangian
of the theory should at least involve the derivatives of
$z^A$, i.e.,  $\partial_a z^A\equiv v^A_a$. The derivatives themselves are not
invariant under transformation (\ref{1}):
\begin{align}
	v'^A_a=\frac{\partial z'^A(y)}{\partial z'^B(y)} v^B_a.
\end{align}
The quantities  $v^A_a$ and the metric of the ambient space
may be used to form the matrix
\begin{align}
	w^{AB}\equiv v^A_a v^B_b \eta^{ab},
\end{align}
which serves as the metric for the directions orthogonal to the surfaces  $\mathcal M$. The upper and lower indices of
the orthogonal space may be raised and lowered using
the matrix. $w^{AB}$ and the inverse matrix $w_{AB}$ Using $w_{AB}$, we then define the projectors to the tangent $\Pi_{ab}$ and normal $\Pi_{\bot ab}$ spaces:
\begin{align}
	\Pi_{\bot ab} \equiv v^A_a v^B_b w_{AB}, \ \Pi_{ab}=\eta_{ab}-\Pi_{\bot ab}.
\end{align}
A direct calculation demonstrates that the projectors
are invariant under transformation  (\ref{1}). However, they
cannot be used for forming any invariant scalar since $\Pi^2=\Pi$ and $\Pi\Pi_\bot$=0, so that at least their derivatives
are required.
Note that the quantity $\Pi_b^{d} \partial_{c} \Pi^a_d$ is not affected by
transformation  (\ref{1}):
\begin{align}
	\Pi_b^{d} \partial_{c} \Pi^a_d \equiv \widehat{b}^a\,_{bc}=\widehat{b}'^a\,_{bc}
\end{align}
Thus the third-rank tensor $\widehat{b}^a\,_{bc}$, is the simplest nontrivial invariant tensor of the considered theory (the
projector derivative reduces to this tensor since $\partial_c \Pi^a_b=\widehat{b}^a\,_{bc}+\widehat{b}^b\,_{ac}$). Therefore, the simplest nontrivial
scalars may be obtained by contracting two such tensors. Taking into account that the tensor $\widehat{b}^a\,_{bc}$ is tangent in the index $b$ and normal in the index $a$, the possible contractions are
\begin{align}
\begin{split}
	&I_1=\widehat{b}^a\,_{bc}\, \widehat{b}_a\,^{bc}, \ I_2=\widehat{b}^a\,_{bc}\, \widehat{b}_a\,^{cb}, \ I_3=\widehat{b}^a\,_{c}\,^c \widehat{b}_a\,^{b}\,_b, \\
	&I_4= \widehat{b}^a\,_{bc} \widehat{b}^c\,_{ba}, \ I_5 = \widehat{b}^a\,_{ba} \widehat{b}^c\,_{bc},
	\end{split}
\end{align}
while all other contractions are identically equal to
zero. The Lagrangian should be scalar under transformations (\ref{1}),  and, therefore, must be a function of the above scalars. If the treatment is restricted to the leading contribution in  $b$, this function may be taken in the
form of a linear combination of the scalars $I_{1-5}$. A calculation shows that the variation of the integral $\int d^{10}y I_1$ is equal to zero up to surface terms.

Invoking the known Gauss formula which interrelates the external and internal geometries and using the expression
\begin{align}
	{b}^a\,_{bc}=\Pi_b^d \widehat{b}^a\,_{dc}
\end{align}
for the second main surface form, it can be shown that
the combination of scalars $I_2-I_3$  represents the scalar
curvature of the  $z^A$ constant-value surface:
\begin{align}
	R={b}^a\,_{bc}\, {b}_a\,^{bc} - {b}^a\,_{c}\,^c {b}_a\,^{b}\,_b.
\end{align}

\section{Comparison with the Einsteinian dynamics}
However, varying the action $\tilde{S}=\int d^{10} y R$ fails to
yield the equations analogous to those of Einstein:
\begin{align}
\frac{\delta \tilde{S}}{\delta z^A} = b_A\,^{ab} G_{ab}+\frac{1}{2}( b_A\,^a\,_a R -  v^a_A \partial_a R )+ (u_a u^a +  \bar{\partial}_a u^a)b_A\,^b\,_b - (u_a  u^b +  \bar{\partial}_a  u^b )b_A\,^a\,_b =0,
\end{align}
where $ \bar{\partial}_a = \Pi^b_a \partial_b$   and $u_a = \widehat{b}^b\,_{a}\,_b$.
In addition to the Einstein tensor, the field equations feature extra terms that cannot be interpreted as part of the Einstein theory. Therefore, the action   $\tilde{S}$ has to be modified so as to accommodate the Einstein limit.

The Einstein limit of the theory may be reached
under the assumption that the  $z^A$ constant-value surfaces do not interact among themselves, i.e., the
action represents an integral over $z$ configurations
from the action of each separate surface,
\begin{align}
	S=\int d^6z S_\mathcal{M}(z),
\end{align}
where, temporarily introducing the surface coordinates, we may write
\begin{align}
	S_\mathcal{M}(z)=\int d^4x \sqrt{-g} \mathcal L(x,z),
\end{align}
so that finally we have
\begin{align}
	S=\int d^6z d^4x \sqrt{-g} \mathcal L(x,z),
\end{align}
where  $\mathcal L(x,z)$ is the surface Lagrangian which may
now be represented by the scalar curvature $R$.

To get rid of the $x^\mu$ one should perform a transition from the curvilinear basis  $\tilde{y}^a = \{x^\mu;z^A\}$ to the Cartesian basis $y^a$. The Jacobian of this transformation $J=\det\dfrac{\partial \tilde{y}^b }{\partial {y}^a }$ has to be calculated. To do so, let us consider two mutually inverse matrices
\begin{align}
	Q^{ab} = \frac{\partial \tilde{y}^a }{\partial {y}^c } \eta^{cd} \frac{\partial \tilde{y}^b }{\partial {y}^d }, \qquad Q^{-1}_{ab} = \frac{\partial {y}^c }{\partial \tilde{y}^a } \eta^{cd} \frac{\partial {y}^d }{\partial \tilde{y}^b }.
\end{align}
Obviously, $	\det Q =   J^2 \det \eta $. The minor of the matrix $Q$
\begin{align}
	\det{Q^{AB}} = \det \left(\frac{\partial \tilde{y}^A }{\partial {y}^c } \eta^{cd} \frac{\partial \tilde{y}^B }{\partial {y}^d }\right)= \det \left(\frac{\partial z^A }{\partial {y}^c } \eta^{cd} \frac{\partial z^B }{\partial {y}^d }\right) = \det w^{AB} = w
\end{align}
and the minor of $Q^{-1}$
\begin{align}
	\det{Q^{-1}_{\mu\nu}} = \det \left(\frac{\partial {y}^c }{\partial \tilde{y}^\mu } \eta^{cd} \frac{\partial {y}^d }{\partial \tilde{y}^\nu }\right)= \det \left(\frac{\partial {y}^c }{\partial x^\mu } \eta^{cd} \frac{\partial {y}^d }{\partial x^\nu }\right) = \det g_{\mu\nu} = g
\end{align}
are known \cite{gantmakher} to satisfy the condition
\begin{align}
	\det{Q^{-1}_{\mu\nu}} = \frac{\det{Q^{AB}}}{\det Q} \Rightarrow g = \frac{w}{J^2 \det \eta} = - \frac{w}{J^2}
\end{align}
whence we conclude that
\begin{align}
	d^6z d^4x \sqrt{-g} \rightarrow d^{10}y |J| \sqrt{-g}  = d^{10}y \frac{\sqrt{|w|}}{\sqrt{-g}} \sqrt{-g}  = d^{10}y \sqrt{|w|}.
\end{align}
The splitting-theory action analogous to that of Einstein-Hilbert is therefore
\begin{align}
	S = \int d^{10}y \sqrt{|w|} R.
\end{align}

This expression features the square root of the determinant of the "metrics"  $w^{AB}$, which is not a scalar
under transformation (\ref{1}) but does not violate the
covariance of equations of motion with respect to this
transformation. This is because the factor  $\sqrt{|w|}$  is a
weight for the contribution of each surface to the overall action and does not affect the equations of motion
as long as there is no interaction among the surfaces.
It is straightforward to demonstrate that no other
function of the determinant $w$ yields the required Einstein limit. Indeed, if we vary the action
\begin{align}
	S=\int d^{10}y f(w)
\end{align}
and require that the equations of motion be analogous
to the equations of motion for a surface,
\begin{align}
	{b}_A\,^{c}\,_c=0,
\end{align}
this leads to the equation $2wf''+f'=0$, which is satisfied only by the functional form $f(w)\sim \sqrt{w}$ .
Therefore, if the theory of gravity is formulated in
terms of a field theory in flat space and the equations of
motion are required to satisfy the Einstein limit, to be
covariant under transformation (\ref{1}) , and to involve the
lowest nontrivial order in $b$ , the action should have the
form\begin{align}
	S=\int d^{10}y \sqrt{w} (R+\alpha I_1 +\beta (I_2+I_3) +\gamma I_4 +\delta I_5).
\end{align}

{\bf
Acknowledgements.} The work of AAS was supported by SPbU grant N 11.38.223.2015.


\begin{thebibliography}{4}
\newcommand{\eprint}[1]{\href{http://arxiv.org/abs/#1}{#1}}
\bibitem{carlip}
 \textit{Carlip S.} // Rept.~Prog.~Phys. 2001. V.~64. P.~885. \eprint{arXiv:gr-qc/0108040}.

\bibitem{gorbrub}
\textit{Gorbunov D.S.}, \textit{Rubakov V.A.} Introduction to the theory if the early Universe. Cosmological perturbations and inflationary theory. Hackensack, New Jersey: World Scientific, 2011.

\bibitem{regge}
\textit{Regge T.}, \textit{Teitelboim C.} General relativity \`a la string: a progress report // {Proceedings of the First Marcel Grossmann Meeting, Trieste, Italy, 1975}.  Ed. Ruffini R. 1977. P.~77. \eprint{arXiv:1612.05256}.

\bibitem{statja25}
\textit{Paston S.A.} // Theor. Math. Phys. 2011. V.~169. No. 2. P.~1611. \eprint{arXiv:1111.1104}.

\bibitem{gantmakher}
\textit{Gantmacher F. R.} The Theory of Matrices. Vol. I. NY: Chelsea, 1959. P. 15.
\end{thebibliography}
\end{document}